# Probing Excitonic Dark States in Single-layer Tungsten Disulfide

Ziliang Ye[1], Ting Cao[2,3], Kevin O'Brien[1], Hanyu Zhu[1], Xiaobo Yin[1], Yuan Wang[1], Steven G. Louie[2,3], Xiang Zhang[1,3]

[1] NSF Nano-scale Science and Engineering Center (NSEC), 3112 Etcheverry Hall, University of California at Berkeley, Berkeley, California 94720, USA.

[2] Department of Physics, University of California at Berkeley, Berkeley, California 94720, USA.

[3] Material Sciences Division, Lawrence Berkeley National Laboratory, 1 Cyclotron Road, Berkeley, California 94720, USA.

**Transition metal dichalcogenide (TMDC) monolayer has recently emerged as an important two-dimensional semiconductor with promising potentials for electronic and optoelectronic devices[1,2]. Unlike semi-metallic graphene, layered TMDC has a sizable band gap[3]. More interestingly, when thinned down to a monolayer, TMDC transforms from an indirect bandgap to a direct bandgap semiconductor[4,5], exhibiting a number of intriguing optical phenomena such as valley selective circular dichroism[6-8], doping dependent charged excitons[9], and strong photocurrent responses[10,11]. In particular, a single layer of TMDC can absorb more than 10% of light, which is unprecedented[12]. However, the fundamental mechanism underlying such a strong light-matter interaction is still under intensive investigation. The observed optical resonance was initially considered to be band-to-band transitions[4,5,13,14]. In contrast, first-principle calculations predicted a much larger quasiparticle band gap size and an optical response that is dominated by excitonic effects[15-17]. In particular, a recent GW plus Bethe-Salpeter equation (GW-BSE) study theoretically predicted a diversity of strongly bound excitons[18]. Here, we report experimental evidence of the exciton dominance mechanism by discovering a series of excitonic dark states in single-layer WS$_2$ using two-photon excitation spectroscopy. In combination with GW-BSE theory, we find the excitons are Wannier excitons in nature but possess extraordinarily large binding energy (~0.7 eV), an order of magnitude larger than that of conventional semiconductors, leading to a quasiparticle band gap of 2.7 eV. These strongly bound exciton states are**

**observed stable even at room temperature. We reveal an exciton series in significant deviation from hydrogen models, with a novel inverse energy dependence on the orbital angular momentum. These excitonic energy levels are experimentally found robust against environmental perturbations. The discovery of excitonic dark states and exceptionally large binding energy not only sheds light on the importance of many-electron effects in this two-dimensional gapped system, but also holds exciting potentials for the device application of TMDC monolayers[19] and their heterostructures[20] in computing, communication and bio-sensing.**

An exciton is a bound state formed by an excited electron-hole pair due to the Coulomb attraction between the two quasiparticles[21]. This bound state often plays an important role in the optical properties of low dimensional materials[22], such as carbon nanotubes[23,24] and quantum dots[25], owing to their strong spatial confinement and reduced screening effect compared to bulk solids. In a 2D gapped system with dipole-allowed interband transitions, the optical absorption spectrum in the non-interacting limit exhibits a step function. Strong electron-hole interaction redshifts a large amount of the spectral weight, resulting in a qualitatively different spectrum with a series of new excitonic levels below the quasiparticle band gap. In quasi-2D quantum wells, the electron-hole interaction is weak[26]. Therefore, by measuring the energy difference between the first excitonic peak and band-edge absorption step, the exciton binding energy can be unambiguously determined, which usually has an energy of 10s of meV and is vulnerable to the environment screening and temperature broadening. However, recent experiments on the single-layer TMDC found no absorption step. Instead, two absorption peaks from spin-orbit splitting were detected[4,5] around the Kohn-Sham band gap energy given by density functional theory (DFT) within the local density approximation. The peaks were initially interpreted as direct band edge transitions. In sharp contrast, more accurate first-principles calculations using the GW method[27,28] predicted a quasiparticle band gap that is larger than the initial experimental reported value by nearly one electron volt[15-18]. This energy gap discrepancy is computed through first-principles GW-BSE theory[29] to be originated from strong excitonic effects[15-18]. It is therefore critical to uncover on firm grounds the underlying physics of the strong light-matter interaction in such a 2D system.

We probed the excitonic effects in TMDC using the two-photon excitation spectroscopy[24,30,31]. At the simplest level, if electron and hole interact through a central attractive Coulomb potential, the electron-hole pair forms a series of excitonic Rydberg-like states with definite parity, similar to the hydrogen model. For $WS_2$, the breaking of rotational and inversion symmetry owing to the crystal structure and the spatial-dependence of screening will modified the energy and symmetry of the states from those of the 2D Rydberg series. However, for exciton states with an electron-hole wavefunction that is large compared to the unit cell size (as shown below for $WS_2$), specific parity may still be assigned to each excitonic state. Incident photons can excite the electronic system from the ground state to one of these excitonic states (Fig. 1(a)). In addition to energy conservation, the selection rule of such a transition depends on the symmetry of the final state: for systems with dipole-allowed interband transitions (which is the case for $WS_2$), one-photon transitions can only reach excitonic states with even parity, while two-photon transitions reach states with odd parity. The two-photon resonances are also known as excitonic dark states as they do not appear in the linear optical spectrum. These dark states are good gauges for excitonic effects, since there is little impurity and bandgap absorption background in the two-photon spectrum. Owing to the direct band gap in TMDC monolayer, we monitor the two-photon absorption induced luminescence (TPL) with a high signal-to-noise ratio. The luminescence results from the radiative recombination of the excitonic ground state, following the rapid non-radiative relaxation from the two-photon excited excitonic dark states to the ground state (Fig. 1(a)). By scanning the excitation laser energy, we obtain a complete two-photon spectrum, assuming the relaxation and emission efficiency are independent of the excitation energy[24].

For our samples, we exfoliate flakes of $WS_2$, which has a higher quantum efficiency than other TMDC monolayers[14], onto a fused quartz substrate from a synthetic $WS_2$ crystal. A typical light emission spectrum is shown in Fig. 1(b), excited by the ultrafast laser (190 fs) at 990 nm (1.25 eV) at 10K. The two peaks observed at 2.0 and 2.04 eV correspond to the exciton and trion emissions from the direct band gap at *K* and *K'* valleys in the Brillouin zone, consistent with the absorption peaks in the reflectance spectrum. The emitted photon energies of both peaks are much higher than those of the excitation photon, and therefore, they can only originate from the two-photon absorption induced luminescence. The two-photon origin of these emissions is further confirmed in the inset to Fig. 1(b). Both the TPL and the SHG signal show a quadratic power dependence, suggesting that the emission is indeed induced by two-photon absorption. The TPL saturates at higher power as a consequence of heating or exciton-exciton annihilation effects[32,33]. For the rest of the experiments, we limit the excitation power to the unsaturated regime. The trion peak amplitude is selected as our TPL signals, since it is stronger than the neutral ground-state exciton emission at 10K.

We observed two important resonances of similar linewidths in the two-photon spectrum, occurring at 2.28 and 2.48 eV, corresponding to two excitonic dark excited states (Fig. 2). The absorption spectrum of $WS_2$ monolayer is plotted for comparison, where the A exciton (the 1$s$ state) and its trion result in two absorption peaks at 2.04 and 2 eV, respectively. Near these one-photon resonances, TPL is negligible, consistent with the 1$s$ nature of these states. On the other hand, no significant one-photon absorption is observed near the excitonic dark states, except for the B exciton (the other 1$s$ state) at 2.45 eV resulted from the spin-orbit splitting in the valence band. Such a complimentary feature reflects the symmetry of the observed excitonic states. Hence, we label the TPL peaks to be the 2$p$ and 3$p$ state of the A exciton series. Accordingly, the 1$s$-2$p$ and 1$s$-3$p$ separations are 0.24 eV and 0.44 eV respectively. The extraordinary large 1$s$-n$p$ (n=2,3) separations suggests that the exciton binding energy, defined as the separation between the 1$s$ exciton ground state and the conduction band edge, is larger than 0.44 eV, which also indicates a significant self-energy contribution to the quasi-particle band gap. Our discovery demonstrates that the previously claimed band-to-band transition mechanism in monolayer TMDC's optical response is qualitatively incorrect, which as we now show is dominated by excitonic states within the band gap, in agreement with the GW-BSE calculation in $MoS_2$ [18]. The real quasiparticle band gap is much larger than previously reported. This finding is expected be general for other TMDC monolayer of similar structures.

We used the *ab initio* GW method[28] to calculate the quasiparticle band structure and the *ab initio* GW-BSE approach[29] to calculate the excitonic states and optical spectrum of a

WS$_2$ monolayer (Fig. 3 A), employing the BerkeleyGW package[34]. The principle and orbital quantum numbers of each exciton state are identified by analyzing the real-space wavefunction's character of the exciton (Fig. 3 B-F). Consistent with the selection rule of one photon absorption for dipole-allowed materials, we find that the "*s*" state is one-photon active or bright, while the other ("*p*" and "*d*") excitons are one-photon inactive or dark. Clearly, the calculated 2*p* and 3*p* states, marked at 2.28 and 2.49 eV, agree well with the experimental results, which confirm our observation of dark excitonic states in WS$_2$ monolayer. The calculated positions of the 1*s* state of the A exciton series (2.04 eV) and B exciton series (2.4 eV) also agree well with the experimental spectrum. As evident from the real-space wavefunctions in Fig. 3 B-F, the excitons in monolayer WS$_2$ have a Wannier nature with their in-plane radii much larger than the unit cell dimension.

In spite of its Wannier character, we found the exciton series in monolayer WS$_2$ deviates significantly from a 2D hydrogen model, which has also been predicted in recent GW-BSE calculations[18,35]. The ratio between 1*s*-2*p* and 1*s*-3*p* separations is 27/32 and 25/27 in 2D and 3D hydrogen models, respectively; neither of which is close to our experimental results or the GW-BSE results (approximately 6/11). In addition, in a hydrogen atom, orbitals with the same principal quantum number are degenerate. However for the WS$_2$ excitons, our calculations show that states in the same shell but of higher orbital angular momentums are at lower energies, i.e., $E_{3d} < E_{3p} < E_{3s}$. Analysis of the theoretical results revealed that these two exotic energy-level behaviors are caused by a strong spatial-dependent dielectric screening: in an atomically thin semiconductor, the screening effect is weaker when the separation between the electron and hole is bigger, which is known as the anti-screening effect in 1D carbon nanotube[36]. Since the wavefunction of excitonic states with higher principal or higher orbital quantum number features a larger nodal structure near the hole (i.e., larger average electron-hole separation), weaker screening at larger separation leads to enhanced Coulomb attraction in the excited states and therefore lowering their excitation energies as compared those of the hydrogen model[36]. Also, because of the degeneracy of the *K* and *K'* valleys in TMDC system, each *s* level has two degenerate states, while each *p* and *d* level has 4 states if perfect rotational symmetry is assumed. All of these features are expected to be quite general for 2D TMDC excitons.

The GW quasiparticle band gap is calculated to be ~ 2.7 eV, labeled by the blue arrow in Fig. 3. Comparing it with the 1$s$ exciton energy found in either our experiment or our GW-BSE calculation, we obtain an exciton binding energy of ~0.7 eV. Such an exceptionally large binding energy is more than ten times larger than the excitons in bulk $WS_2$[3] as well as other traditional bulk semiconductors such as Si and GaAs[21] and comparable to those in carbon nanotubes[23,24], resulted from the combined effects of reduced dimensionality, relatively large effective masses and weak dielectric screening, which renders the excitons observable even at room temperature.

The highly localized exciton wavefunction in the out-of-plane direction, on one hand, indicates a tightly bound exciton which maybe immune to environment perturbation. On the other hand, 2D materials are usually sensitive to external dielectric screening as all the atoms are near the surface. To explore such an effect, we measure the two-photon spectrum of monolayer $WS_2$ with different dielectric capping layers including water, immersion oil and aluminum oxide with their average dielectric constant at optical frequency ranging from 1.7 to 2.5.

In all capped samples, we observed the 2$p$ and 3$p$ resonances even at room temperature, as expected from the large exciton binding energy. (Fig. 4(a)) We find no significant shift in the excitation energy of either the $s$ or the $p$ states with different capping layers, except for an overall temperature related redshift (0.04 eV) and linewidth broadening compared with measurement at 10k (Fig. 2). For the lowest bound excitonic state, the insensitivity of the emitted photon energy to external dielectric screening can be understood as the environmental screening's opposite effects to the electron self-energy and the exciton binding energy. The same argument may apply to the lower-energy excited states. Nevertheless, it is interesting that, with different capping, the 1$s$-2$p$ and 1$s$-3$p$ energy differences remain roughly unchanged, ~0.2 and 0.5 eV, respectively. This robustness indicates the measured excitation energies for the 2$p$ and 3$p$ states are intrinsic to the monolayer, thus agreeing well with those from *ab-initio* GW-BSE calculation for the vacuum condition. Together with the TPL signal, SHG is also observed as a slanted straight line in the excitation-emission spectra (Fig. 4(b)). At room temperature, the exciton-trion separation is no longer distinguishable, but the 2$p$ and 3$p$ absorption peaks remain prominent. A SHG resonance occurs as the TPL and SHG line cross each other, known as the exciton enhanced SHG effect[37].

SHG signals are originated from the broken inversion symmetry within the $WS_2$ monolayer[37,38]. As discussed above, strictly speaking, parity is not a good quantum number for any inversion-symmetry-broken system, which may explain why TPL resonances are superimposed on a plateau background. But similar to the approximation that GaAs quantum well can be treated as of $D_{4h}$ group with inversion symmetry if only transitions near the Brillion zone center are considered[39], we can approximately restore inversion symmetry in TMDC monolayer by only considering transitions within a single valley and ignoring the spin, which is valid in the experiment and so is the TPL selection rule. The TPL selection rule is different from the valley selection rule of circularly polarized photon absorption[6-8], though the combination of them deserves further exploration.

In summary, we experimentally reveal the two-dimensional excitonic dark states and strong excitonic dominance mechanism in $WS_2$ monolayer. The $WS_2$ excitons are observable even at room temperature, and are Wannier-like in nature but with an exceptional large binding energy of 0.7 eV for the 1$s$ state which significantly changes the optical gap from the quasiparticle band gap. The novel exciton series is found in substantial deviation from hydrogen models, with the excitation energy of the low-energy exciton states robust to environment perturbation. These phenomenal physical properties unveil an intense many-electron effect in this class of 2D gapped systems. The determined band gap size allows us to accurately design heterostructures between a TMDC monolayer and other materials. Discovery of extraordinarily strong excitons in TMDC sets an important foundation for exploiting the unusual light-matter interactions from strong many-electron effects, as well as the emerging 2D electronic and optoelectronic applications.

**Methods:**

**Sample preparations:**

WS$_2$ samples are directly exfoliated onto fused quartz substrates from a synthesized crystal (2d Semiconductors Inc.). The exfoliated monolayer flake is normally a few micrometers in size and characterized by tools such as AFM, micro Raman and photoluminescence. The solid-state capping is 50 nm thick Al$_2$O$_3$, coated with the Atomic Layer Deposition (ALD) technique. The liquid capping is prepared by wetting the sample with deionized water or immersion oil (Zeiss, Immersol 518 F).

**Two-photon excitation spectroscopy:**

The excitation spectroscopy is carried out with an optical parametric oscillator (Newport, Inspire HF 100) pumped by a mode-locked Ti:sapphire oscillator. The laser pulse width is about 190 (±20) fs and repetition rate is 80 MHz. The low temperature experiment is operated in a continuous-flow liquid Helium cryostat equipped with a long working distance 50x objective of a 0.55 NA and the room temperature data is collected by a 100x objective of a 0.9 NA or a 1.4 NA 63x oil immersion objective. The emission signal is detected in the back scattering configuration and analyzed by a cooled CCD spectrometer. The transmissivity of the optical system is carefully calibrated to evaluate the absolute power level at the focusing plane. The emission spectra are normalized to the square of the focused power, as the excitation is limited to the unsaturated regime. The laser pulse width is measured by a home-built autocorrelator at the focus throughout the scanning range. The micro reflectivity spectrum is taken with a focused supercontinuum laser (Fianium, SC450).

**First-principles calculations:**

Density functional calculations are performed using the local density approximation (LDA) implemented in the Quantum Espresso package[40]. The GW and GW+BSE calculations are performed with the BerkeleyGW package[34]. The dielectric matrix is constructed with a cutoff energy of 476 eV. The dielectric matrix and the self-energy are calculated on a 18x18x1 k-grid. The quasi-particle band gap is converged to within 0.05 eV. In the calculation of optical absorption spectra, the quasi-particle band structure and electron-hole interaction kernal are interpolated onto a 81x81x1 fine k-grid, with the 1s exciton binding energy converged to within 0.05 eV. The spin-orbit coupling is included perturbatively.

Figure 1:

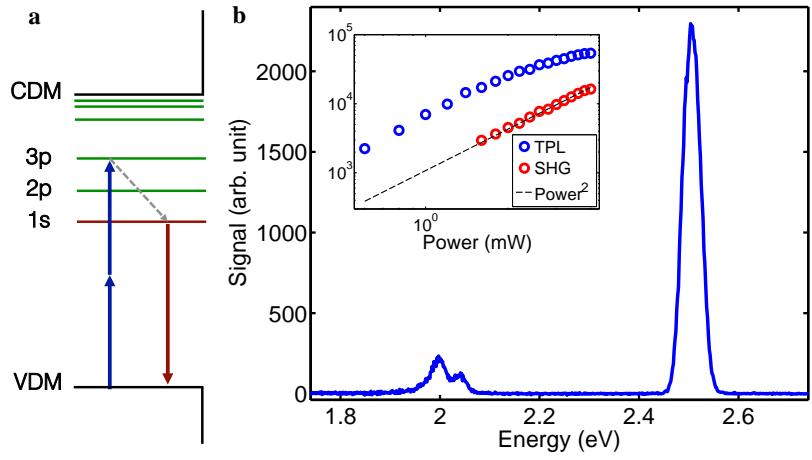

Figure 2:

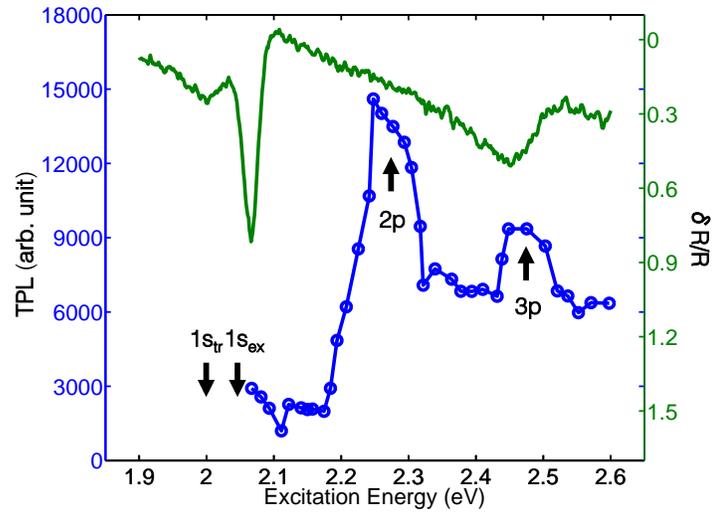

Figure 3:

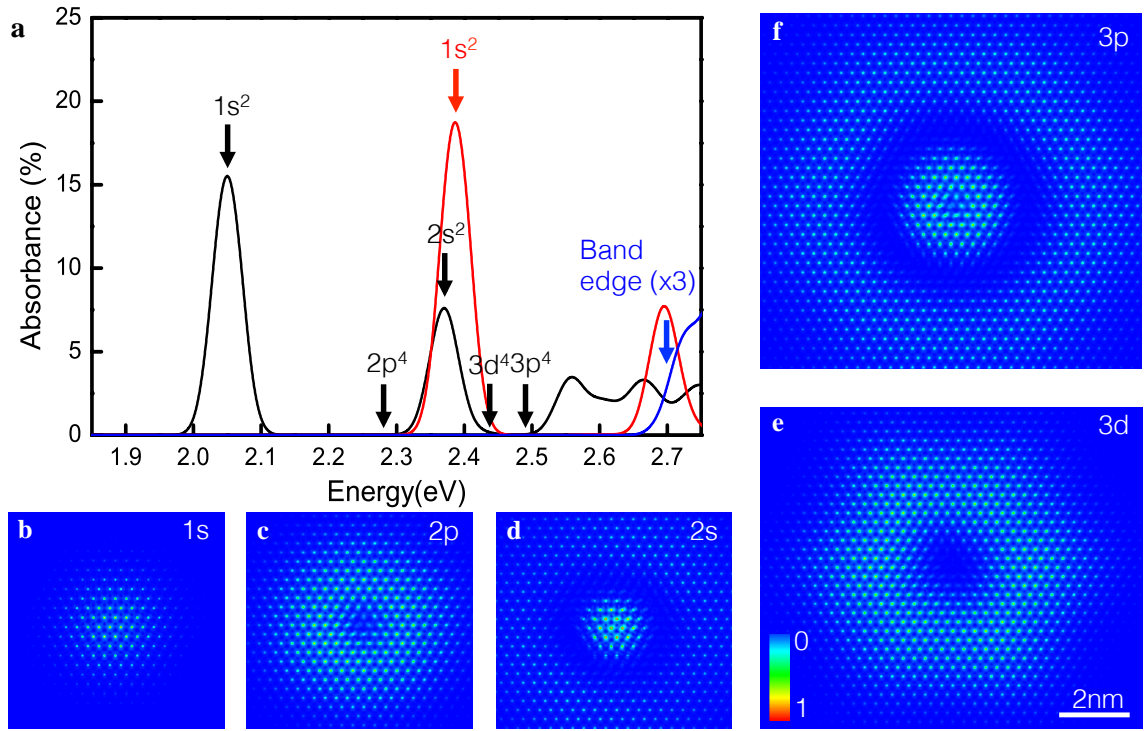

Figure 4:

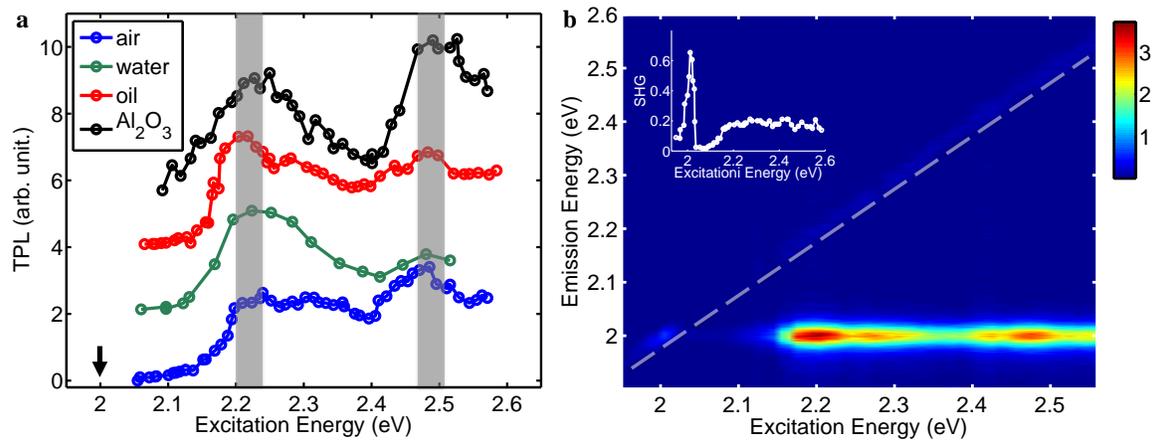

**Figure Captions:**

**Figure 1: Probing the dark exciton states in a single-layer $WS_2$ by two photon luminescence (TPL)**

a. Schematic of the TPL process in a single-layer $WS_2$. Under the two-photon excitation, electron transitions to one of the excitonic dark states with odd parity (double blue arrow). Following the excitation, the exciton experiences a fast relaxation to the excitonic ground state (grey arrow) and emits a photon (red arrow). The two-photon selection rule exclusively eliminates the one-photon transition background and reveals the excitonic excited states. *s* (red) and *p* (green) characters are labeled according to the excitonic envelope wavefunction character.

b. Measured $WS_2$ emission spectrum excited by an ultrafast laser pulse at 10K. The peaks at 2.04 eV and 2 eV are the A exciton (1*s* state) and its trion peak, respectively. The lower-energy peak is stronger than the higher-energy one due to the exciton-trion equilibrium reached during the emission stage at low temperature. The excitation pulse is at 1.25 eV with a pulse width of about 190 (±20) fs, which results in the 2.5 eV peak as the SHG signal. In the inset, the power dependences of SHG and TPL signal are plotted. At a low excitation level, both of them exhibit quadratic power dependence, confirming the two-photon absorption nature of the luminescence, until the TPL signal saturates at a high excitation level. In the low temperature experiment, the TPL signal represents the peak amplitude of the trion peak.

**Figure 2: Extraordinarily strong excitonic effect in monolayer $WS_2$**

Two-photon absorption (blue) and one-photon absorption (green) spectra are measured in a single-layer $WS_2$ at 10K. In the two-photon absorption spectrum, 2*p* and 3*p* resonances are observed at 2.28 eV and 2.48 eV, on top of a plateau background. For comparison, the one-photon absorption spectrum, measured as the relative reflectance signal, exhibits no corresponding features except a B exciton (1*s*) related absorption resonance at 2.45 eV. Additionally, the A exciton and trion (1*s*) absorption peaks are detected consistently with the TPL emission peaks (Fig. 1 b), with a 20 meV Stoke shift, and are marked at 2.04 and

2 eV, respectively. The energy difference between the A exciton 1$s$ state emission peak and the 3$p$ state absorption peak is 0.44 eV which yields the lower bound for the exciton binding energy in monolayer WS$_2$. This binding energy is extraordinarily large for a Wannier exciton, and implies a dominating excitonic mechanism for the intense light-matter interaction in 2D TMDC. The total excitation scan is achieved by tuning an output beam of an optical parametric oscillator over a 600 meV span, with a scanning resolution about 15 meV. (see Methods)

**Figure 3: One-photon absorption spectra and real-space exciton wavefunctions from ab initio GW-BSE calculation.**

a. The optical absorption of A (black) and B (red) exciton series. The blue curve is the optical absorption spectrum without considering electron-hole interaction, where the quasiparticle band gap is about 2.7 eV (blue arrow). With electron-hole interaction, the excitonic states of A and B exciton series are calculated (b-f) and labeled by black and red arrows, respectively, up to 2.5 eV. The computed 1$s$, 2$p$ and 3$p$ states of the A exciton are at 2.05 eV, 2.28 eV and 2.49 eV, respectively, and are in excellent agreement with the experimental measurements. Although the orbital notation of a 2D hydrogen atom is adopted to label the exciton states, the excitonic series significantly deviates from any hydrogenic series, as discussed in the main text. The degeneracy labels in the superscript include both the degeneracy of valleys and orbital angular momentum.

b-f. The real-space plots are modulus squared of the exciton wavefunction projected onto the WS$_2$ plane, with the hole position fixed near a Mo atom at the center of the plot. These wavefunctions share similar in-plane nodal structures with the excited states in a hydrogen atom, and therefore enables the eigenstate being labeled with a principal and an orbital quantum number. The Wannier nature of the excitons is clear with the radii much larger than the unit cell.

**Figure 4: Robust excitonic energy levels to the dielectric environment and temperature effects**

a. Room-temperature two-photon spectra of single-layer $WS_2$ with different top capping layers that tune the dielectric environment immediately adjacent to the atomic layer. The curves respectively represent the uncapped ($\varepsilon_{ave}$=1.625), water capped ($\varepsilon_{ave}$=1.97), immersion-oil capped ($\varepsilon_{ave}$=2.25), and $Al_2O_3$ capped ($\varepsilon_{ave}$=2.57) samples, and each curve is adjusted to a similar vertical scale and shifted for better visualization. The emission peak is at 2 eV, marked by the black arrow. Evidently, the 2*p* and 3*p* peak positions remain roughly unchanged within the experimental error, marked by the grey bands at 2.22 (±0.02) eV and 2.49 (±0.02) eV, respectively. Therefore, the 1*s*-n*p* (n=2,3) separation is approximately the same as the low-temperature uncapped result (Fig. 2), suggesting the excitation energy of the low-energy exciton levels are relatively insensitive to dielectric environmental and temperature perturbations, as discussed in the main text.

b. Measured emission spectra at different excitation energies of an immersion-oil capped $WS_2$ monolayer at room temperature. The horizontal line signal is the TPL emission, with two hotspots along the line corresponding to the 2*p* and 3*p* two-photon absorption peaks. The SHG signal due to the broken inversion symmetry in the monolayer is observed (along the dashed line as an eye guide). At the intersection between the SHG and TPL line, the SHG signal experiences an excitonic enhancement from the A exciton 1*s* state (inset).